# Spontaneous symmetry breaking
# in the quantum sine-Gordon model

S.G. Chung

Department of Physics, Western Michigan University, Kalamazoo, MI 49008-5151

Abstract

The spontaneous symmetry breaking in the quantum sine-Gordon model is demonstrated by a density matrix renormalization group. A phase diagram in the coupling constant - system size plane is obtained.

The sine-Gordon (SG) model has been basically understood, i.e., the Bethe Ansatz (BA) solution[1] and statistical mechanics,[2] in the attractive regime $\beta^2 < 4\pi$. The repulsive regime $4\pi < \beta^2 < \infty$, however, is still open. While a considerable effort has been made, nonexactness of bosonization[3] which connects fermions and bosons limits true SG thoeries only to those of quantum inverse scattering method[4] and perturbative renormalization group studies[5,6]. In short, a unified theory of the SG model which gives the exact soliton spectrum at $4\pi < \beta^2 < 8\pi$, and the massless phase at $8\pi < \beta^2$ is yet to be constructed. A recent work by Kehrein[7] based on Wegner's flow equation method is a good progress in the SG theory. What we know for sure about the SG model is that it undergoes a Berezinskii-Kosterlitz-Thouless (BKT) transition at $\beta^2 = 8\pi$ in the small mass parameter limit. The BKT transition, however, suffers a strong finite-size effect [8] and physical quantities of finite condensed matter systems critically depend on the system size.

The purpose of this paper is to precisely analyze a finite, lattice SG Hamiltonian,

$$H = \sum_{i=1}^{L} \left\{ -\frac{\beta^2}{2} \frac{d^2}{d\phi_i^2} + \frac{1}{2\beta^2} (\phi_i - \phi_{i+1})^2 + \frac{1}{\beta^2} (1 + \cos \phi_i) \right\} \quad (1)$$

where $\phi_i$ is the field variable at the lattice site i and L is the system size. Note that (1), as a mechanical analog, describes a system of torsion-coupled quantum pendula under gravity. Using a density matrix renormalization group (DMRG)[9], we demonstrate the spontaneous symmetry breaking (SSB) in the finite, lattice SG model. We draw a phase diagram in the $\beta^2$ - L plane, a critical line separating the SSB ground state and unbroken one.

To calculate the ground state and the first excited state as a function of the system size L, we follow the standard DMRG procedure. We use the infinite algorithm, open boundary condition, and the ground state target. See Ref. 10 for details of DMRG applied to boson systems. In particular, we use the 3-site algorithm ( add a site in each DMRG step) rather than 4-site algorithm. A special note here is that we truncate the phase space to [$-4\pi, 4\pi$] which makes the massless phase at $\beta^2 > 8\pi$ massive. Nevertheless, as we will see below, we can demonstrate the SSB.

Fig. 1 shows the probability distribution of the phase (position of pendulum in mechanical analog) at the center site in the ground state for $\beta^2 = 13$. The probability distributions at different sites differ only a few % at the edges. Due to the phase space truncation to [$-4\pi, 4\pi$], the translational symmetry is somewhat broken from the outset, and the symmetry unbroken state at $L = 7$ is delocalized over the two potential minima at $-\pi$ and $\pi$. With the increase of the system size, the distribution becomes asymmetric and eventually localized near the potential well at $-\pi$. At the same time, the first excited state shows similar localization but at the other potential minimum at $\pi$. The asymmetry of $-\pi$ ground state and $\pi$ 1-st excited state but not the other way around must be due to a numerical noise. $L = 43$, the two states are almost degenerate, the energy difference $\sim 10^{-5}$, showing the SSB and the associated ground state degeneracy. Fig. 2 shows the phase averages at the center site for the lowest 2 states as functions of the system size L. After $L = 43$, the first excited state suddenly acquires a mass, indicating that the $+\pi$ ground state localized at the potential well at $+\pi$ is no more accessible from the $-\pi$ ground state, and the excited state thereafter is due to a local deformation of the $-\pi$ ground state which must be a topologically neutral soliton-antisoliton pair creation. Fig. 3 shows the phase average, now different from site to site, vs the lattice site in the first excited state at $L = 67$.

We have repeated the calculation varying the coupling constant $\beta^2$. With the decrease of $\beta^2$, the SSB occurs for shorter system sizes and more abruptly. Fig. 4 shows a phase diagram in $\beta^2$ -

L plane with a critical line separating the broken symmetry ground state and the unbroken one.

Finally, an interesting question is if the SSB and the BKT transition are identical. The answer to this question is YES, but this issue will be discussed elsewhere.[11] An important application of the present work in condensed matter systems will be to a small arrays of Josephson junctions. The spontaneous symmetry breaking will then describe the superconductor-insulator transition.

I thank David Kaup for his correspondence. This work was partially supported by NSF under DMR980009N and DMR990002N and utilized the SGI/CRAY Origin2000 at the National Center for Supercomputing Applications, University of Illinois at Urbana-Champaign.

**Figure Captions**

Fig. 1    The probability distribution of the phase at the center site in the ground state for the system sizes L = 7(cross), 37(triangle), 43(circle) and 61(square). $\beta^2 = 13$.

Fig. 2    The phase average vs the system size for the lowest 2 states for $\beta^2 = 13$. The phase-average split to $\pm\pi$ and energy degeneracy indicate the spontaneous symmetry breaking.

Fig. 3    Phase averages vs the lattice site in the first excited state for $\beta^2 = 13$. Note that in comparison, for ground state, the phase averages are simply at constant -$\pi$.

Fig. 4    The phase diagram in $\beta^2$ - L plane.

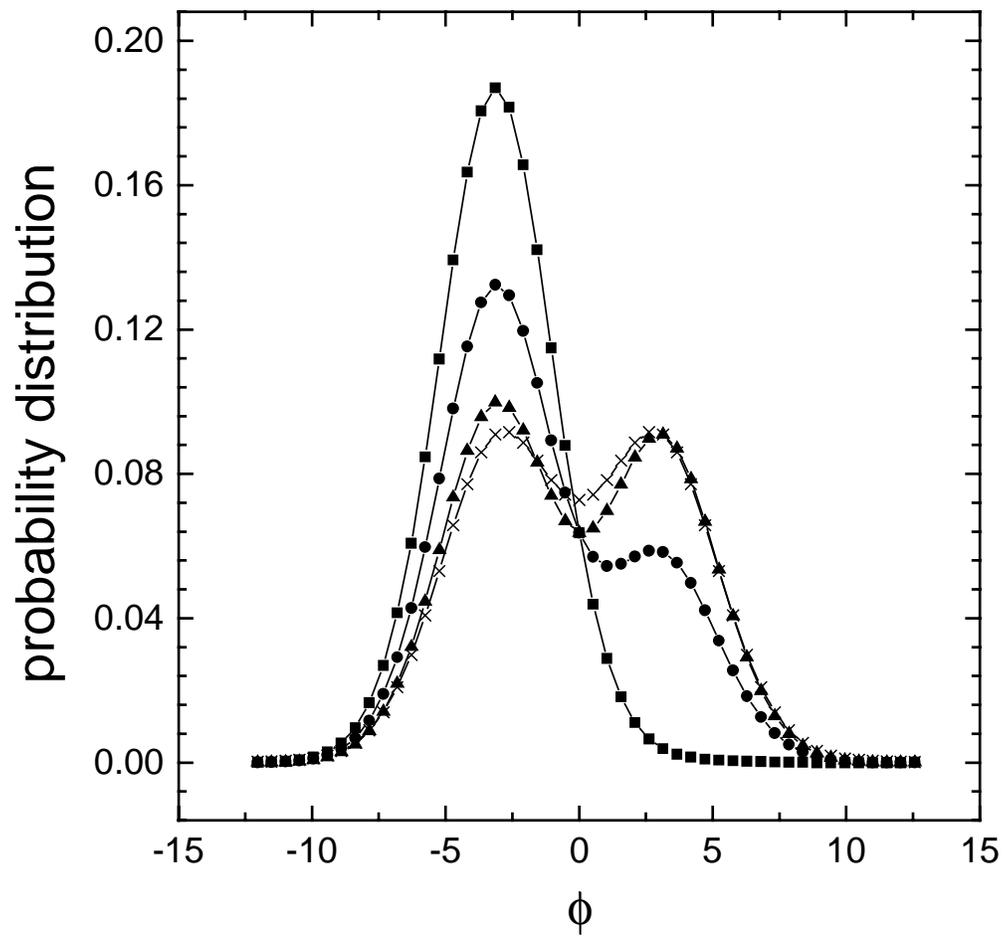

Fig 1

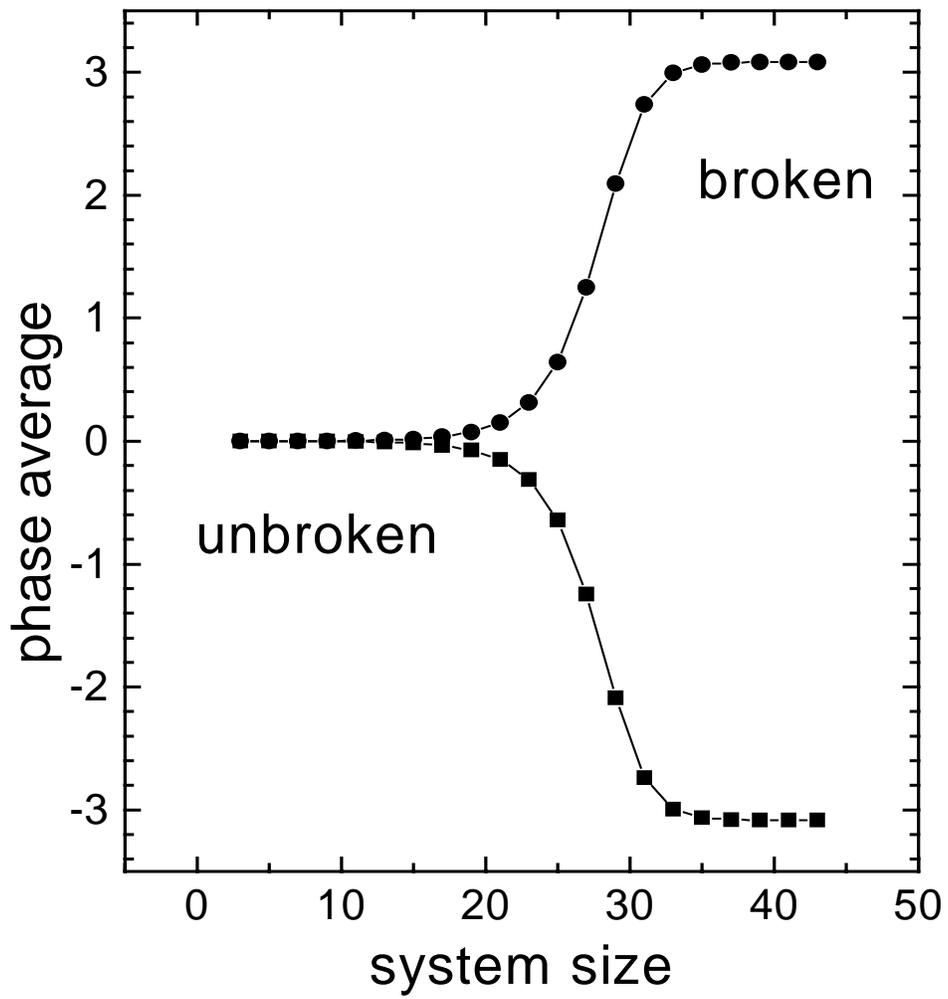

Fig 2

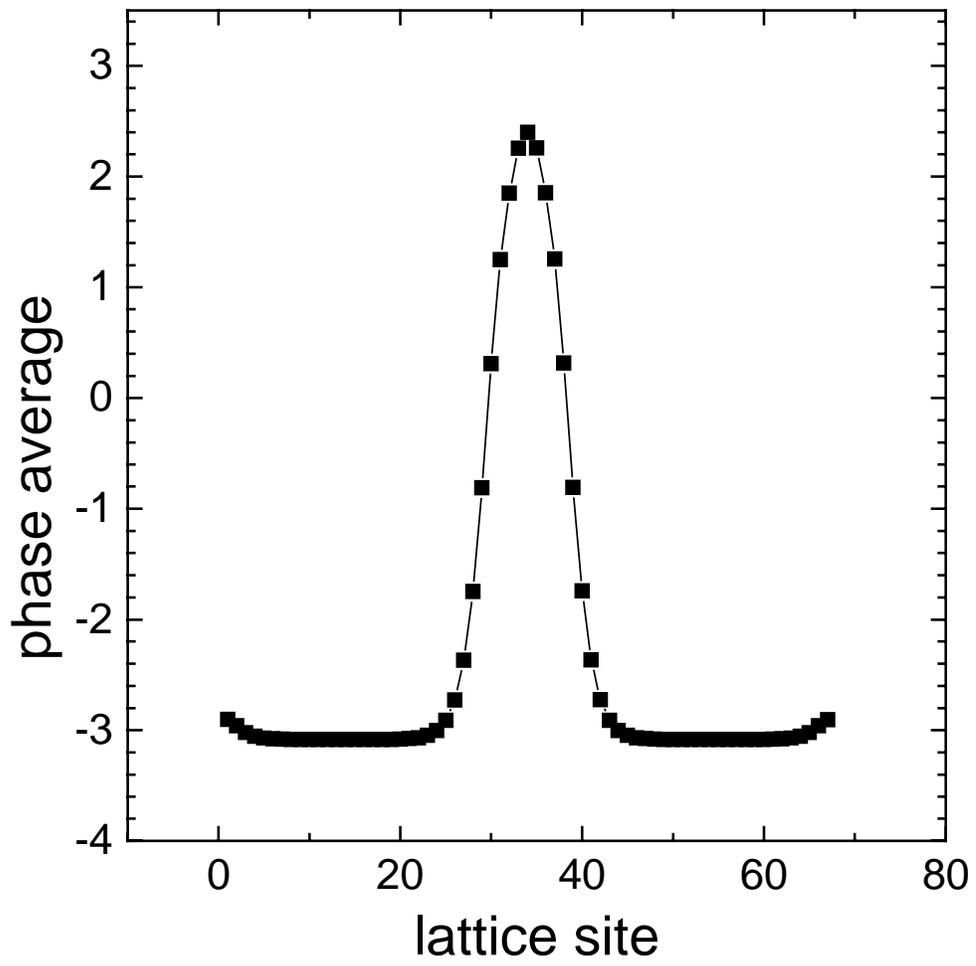

Fig 3

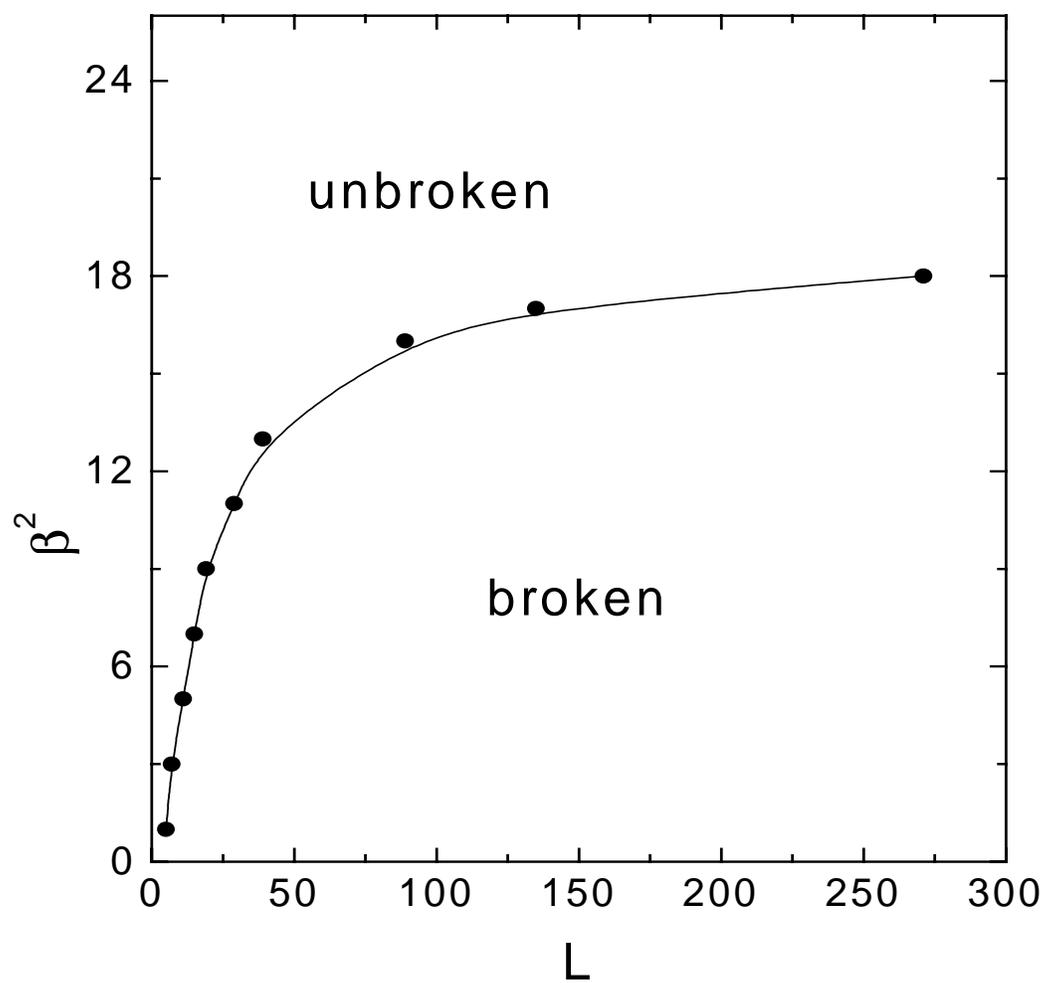

Fig 4